\documentclass[runningheads]{svmult}
\usepackage{graphicx}  

\begin{document}

\def\sin{{\rm sin}}
\def\cos{{\rm cos}}

\title*{Recent Experimental Tests of \\ Special Relativity}

\author{Peter Wolf\inst{1,2}
\and S\'{e}bastien Bize\inst{1}
\and Michael E. Tobar\inst{3}
\and Fr\'ederic Chapelet\inst{1}
\and Andr\'{e} Clairon\inst{1}
\and Andr\'{e} N. Luiten\inst{3}
\and Giorgio Santarelli\inst{1}}

\authorrunning{Peter Wolf et al.}

\institute{BNM-SYRTE, Observatoire de Paris, 61 avenue de l'Observatoire, 75014 Paris, France
\and Bureau International des Poids et Mesures, Pavillon de Breteuil, 92312 S\`evres Cedex, France
\and University of Western Australia, School of Physics, Nedlands 6907 WA, Australia}

\maketitle

\begin{abstract}
We review our recent Michelson-Morley (MM) and Kennedy-Thorndike (KT) experiment, which tests Lorentz invariance in the photon sector, and report first results of our ongoing atomic clock test of Lorentz invariance in the matter sector. 

The MM-KT experiment compares a cryogenic microwave resonator to a hydrogen maser, and has set the most stringent limit on a number of parameters in alternative theories to special relativity. In the Robertson-Mansouri-Sexl (RMS) framework our experiment constrains $1/2 - \beta_{MS} + \delta_{MS} = (1.2 \pm 2.2) \times 10^{-9}$ and $\beta_{MS} - \alpha_{MS} - 1 = (1.6 \pm 3.0) \times 10^{-7}$, which is of the same order as the best results from other experiments for the former and represents a 70 fold improvement for the latter. In the photon sector of the general Lorentz violating standard model extension (SME), our experiment limits 4 components of the $\tilde{\kappa}_{e-}$ parameter to a few parts in $10^{-15}$ and the three components of $\tilde{\kappa}_{o+}$ to a few parts in $10^{-11}$. This corresponds to an improvement by up to a factor 10 on best previous limits.

We also report first results of a test of Lorentz invariance in the SME matter sector, using Zeeman transitions in a laser cooled Cs atomic fountain clock. We describe the experiment together with the theoretical model and analysis. Recent experimental results are presented and analyzed, including statistical uncertainties and a brief discussion of systematic effects. Based on these results, we give a first estimate of components of the $\tilde{c}^p$ parameters of the SME matter sector. A full analysis of systematic effects is still in progress, and will be the subject of a future publication together with our final results. If confirmed, the present limits would correspond to first ever measurements of some $\tilde{c}^p$ components, and improvements by 11 and 14 orders of magnitude on others. 

\end{abstract}

\section{Introduction} \label{intro}

One hundred years after Einstein's first paper \cite{Einstein1905} special relativity is still standing up to all experimental tests and verifications. Over the last century a large number of such tests have provided what is certainly one of the most solid experimental bases of any present fundamental theory of physics. As a consequence special relativity is today underpinning all of present day physics, ranging from the standard model of particle physics (including nuclear and atomic physics) to general relativity and astronomy. That fact continues to push experimentalists to search for new experiments, or improve on previous ones, in order to uncover a possible violation of special relativity, as that would most certainly lead the way to a new conception of physics and of the universe surrounding us. Additional incentive for such tests comes from unification theories (e.g. string theories, loop quantum gravity), some of which \cite{KostoSam,Damour1,Gambini} suggest a violation of special relativity at some, a priori unknown, level. Given the strong theoretical motivation for such theories, but the lack of experimental data that would allow a more rigorous selection among the candidate theories and the parameter space of each class of such theories, any experimental results that could aid the theoretical efforts are certainly welcome.

The fundamental hypothesis of special relativity is what Einstein termed the "principle of relativity" \cite{Einstein1905}, or in more modern terms Local Lorentz Invariance (LLI) \cite{Will}. Loosely stated, LLI postulates that the outcome of any local test experiment is independent of the velocity of the freely falling apparatus. LLI can be viewed as a constituent part of the Einstein Equivalence Principle which is fundamental to general relativity and all metric theories of gravitation \cite{Will}. The experiments presented in this paper test some aspect of LLI, as characterized in Lorentz violating theoretical frameworks like the ones briefly described in section \ref{theo}.

We review and present two of our recent and ongoing experiments \cite{Wolf2003,WolfGRG,Wolf2004} that test different aspects of LLI, analyzing and describing their outcome in two theoretical frameworks, the kinematical test theory of Robertson, Mansouri and Sexl (RMS) \cite{Robertson,MaS} and the Lorentz violating extension of the standard model (SME) \cite{Kosto1}. These experiments, a Michelson-Morley and Kennedy-Thorndike test (section \ref{Molly}), and an ongoing atomic clock test in the SME matter sector (section \ref{FO2}), are among the most precise LLI tests at present.

The vast majority of modern experiments that test LLI rely essentially on the stability of atomic clocks and macroscopic resonators, therefore improvements in oscillator technology have gone hand in hand with improved tests of LLI. The experiments presented here are no exception. All of them employ clocks and resonators developed and used primarily for other purposes (national and international time scales, frequency calibration, etc.) but adapted for tests of LLI.

\section{Theoretical frameworks} \label{theo}

Numerous test theories that allow the modeling and interpretation of experiments that test LLI have been developed. Kinematical frameworks \cite{Robertson,MaS} postulate a simple parametrisation of the Lorentz transformations with experiments setting limits on the deviation of those parameters from their special relativistic values. A more fundamental approach is offered by theories that parametrise the coupling between gravitational and non-gravitational fields (TH$\epsilon\mu$ \cite{LightLee,Will,Blanchet} or $\chi$g \cite{Ni} formalisms) which allow the comparison of experiments that test different aspects of the EEP. Finally, formalisms motivated by unification theories \cite{Damour1,Damour2,Kosto1} have the advantage of opening the way to experimental investigations in the domain of the unification of gravity with the other fundamental forces of nature. In this work we restrict ourselves to two theoretical frameworks, the kinematical framework developed by Robertson, Mansouri and Sexl (RMS) and the more recent standard model extension (SME) of Kostelck\'y and co-workers.

By construction, kinematical frameworks do not allow for any
dynamical effects on the measurement apparatus. This implies that
in all inertial frames two clocks of different nature (e.g. based
on different atomic species) run at the same relative rate, and two
length standards made of different materials keep their relative
lengths. Coordinates are defined by the clocks and length
standards, and only the transformations between those coordinate
systems are modified. In general this leads to observable effects
on light propagation in moving frames but, by definition, to no
observable effects on clocks and length standards. In particular,
no attempt is made at explaining the underlying physics (e.g.
modified Maxwell and/or Dirac equations) that could lead to
Lorentz violating light propagation but leave e.g. atomic energy
levels unchanged. On the other hand dynamical frameworks (e.g. the
TH$\epsilon\mu$ formalism or the SME) in general use a modified
general Lagrangian that leads to modified Maxwell and Dirac
equations and hence to Lorentz violating light propagation and
atomic properties, which is why they are considered more
fundamental and more complete than the kinematical frameworks.
Furthermore, as shown in \cite{KM}, the SME is kept sufficiently
general to, in fact, encompass the kinematical frameworks and some
other dynamical frameworks (in particular the TH$\epsilon\mu$
formalism) as special cases, although there are no simple and
direct relationships between the respective parameters.

\subsection{The Robertson, Mansouri \& Sexl framework} \label{RMStheo}

Kinematical frameworks for the description of Lorentz violation
have been pioneered by Robertson \cite{Robertson} and further
refined by Mansouri and Sexl \cite{MaS} and others. Fundamentally
the different versions of these frameworks are equivalent, and
relations between their parameters are readily obtained. As
mentioned above these frameworks postulate generalized
transformations between a preferred frame candidate $\Sigma(T,{\bf
X})$ and a moving frame $S(t,{\bf x})$ where it is assumed
that in both frames coordinates are realized by identical
standards. The transformations of \cite{MaS} (in
differential form) for the case where the velocity of $S$ as
measured in $\Sigma$ is along the positive X-axis, and assuming
Einstein synchronization in $S$ (in all of the following the choice of synchronization convention plays no role) are

\begin{equation}
dT = {1\over a}\left(dt+{vdx\over c^2}\right); dX = {dx\over
b}+{v\over a}\left(dt+{vdx\over c^2}\right); dY = {dy\over d}; dZ
= {dz\over d} \label{MStransf}
\end{equation}
with $c$ the velocity of light in vacuum in $\Sigma$, and ${\bf v}$ the velocity of $S$ in $\Sigma$. In special relativity $\alpha_{\mathrm{MS}} =
-1/2; \beta_{\mathrm{MS}} = 1/2; \delta_{\mathrm{MS}} = 0$ and
(\ref{MStransf}) reduces to the usual Lorentz transformations.
Generally, the best candidate for $\Sigma$ is taken to be the
frame of the cosmic microwave background (CMB) \cite{Fixsen,Lubin}
with the velocity of the solar system in that frame taken as
$v_\odot \approx 377$ km/s, decl. $\approx -6.4 ^\circ $, $RA
\approx 11.2$h.

Michelson-Morley type experiments \cite{MM} determine the
coefficient $P_{MM} = (1/2-\beta_{\mathrm{MS}}
+\delta_{\mathrm{MS}})$ of the direction dependent term. For many
years the most stringent limit on that parameter was $|P_{MM}|
\leq 5 \times 10^{-9}$ determined over 23 years ago in an
outstanding experiment \cite{Brillet}. Our experiment \cite{WolfGRG} confirms
that result with roughly equivalent uncertainty $(2.2 \times
10^{-9})$. Recently an improvement to $|P_{MM}| \leq 1.5 \times
10^{-9}$ has been reported \cite{Muller}. Kennedy-Thorndike
experiments \cite{KT} measure the coefficient
$P_{KT} = (\beta_{\mathrm{MS}} -\alpha_{\mathrm{MS}} -1)$ of the
velocity dependent term. The most stringent limit \cite{Schiller}
on $|P_{KT}|$ has been recently improved from \cite{Hils} by a
factor 3 to $|P_{KT}| \leq 2.1 \times 10^{-5}$. Our experiment \cite{WolfGRG} improves this
result by a factor of 70 to $|P_{KT}| \leq 3.0 \times 10^{-7}$.
Finally Ives-Stilwell experiments \cite{IS} measure $\alpha_{\mathrm{MS}}$. The most stringent result comes from the recent experiment of \cite{Saathoff} which improves by a factor 4 our 1997 results \cite{WP}, limiting $|\alpha_{\mathrm{MS}} + 1/2|$ to $\leq 2.2 \times
10^{-7}$. The three types of experiments taken together then
completely characterize any deviation from Lorentz invariance in
this particular test theory, with present limits summarized in
table \ref{MStab} (but note that table \ref{MStab} does not include new limits reported in these proceedings).

\begin{table}
\caption{Present limits on Lorentz violating parameters in the
framework of \cite{MaS}, not including new limits reported in these proceedings.}
\begin{center}
\renewcommand{\arraystretch}{1.4}
\setlength\tabcolsep{5pt}
\begin{tabular}{cccc}
\hline\hline\noalign{\smallskip}
Reference & $\alpha_{\mathrm{MS}} + 1/2$ & $1/2-\beta_{\mathrm{MS}} +\delta_{\mathrm{MS}}$ & $\beta_{\mathrm{MS}} -\alpha_{\mathrm{MS}} -1$ \\
\noalign{\smallskip} \hline \noalign{\smallskip}
Saathoff et al. 2003 \cite{Saathoff} & $\leq 2.2 \times 10^{-7}$ & - & -  \\
M\"uller et al. 2003 \cite{Muller} & - & $(2.2 \pm 1.5)\times 10^{-9}$ & - \\
Braxmaier et al. 2002\cite{Schiller} & - & - & $(1.9 \pm 2.1)\times 10^{-5}$ \\
Wolf and Petit 1997 \cite{WP} & $\leq 8 \times 10^{-7}$ & - & -  \\
Wolf et al. 2003 \cite{WolfGRG} & - & $(1.2 \pm 2.2)\times 10^{-9}$ & $(1.6 \pm 3.0)\times 10^{-7}$ \\
\hline\hline
\end{tabular}
\end{center}
\label{MStab}
\end{table}

\subsection{The Standard Model Extension} \label{SMEtheo}

The general Lorentz violating Standard Model Extension (SME) was developed relatively recently by Kosteleck\'y and co-workers \cite{Kosto1}, motivated initially by possible Lorentz violating phenomenological effects of string theory \cite{KostoSam}. It consists of a parametrised version of the standard model Lagrangian that includes all Lorentz violating terms that can be formed from known fields, and includes (in its most recent version \cite{KostoGrav}) gravity. 

The fundamental theory of the SME as applied to electrodynamics is laid out in \cite{KM} and summarized below. We use that approach to model the MM and KT experiments in section \ref{MollySME}. For the discussion of the atomic clock experiment of section \ref{FO2} the SME matter sector is relevant. Its application to atomic physics, and in particular atomic clock experiments, is laid out in \cite{KL,Bluhm} and summarized below. 

Generally, the SME characterizes a potential Lorentz violation using a number of parameters that are all zero in standard (non Lorentz violating) physics. These parameters are frame dependent and consequently vary as a function of the coordinate system chosen to analyze a given experiment. In principle they may be constant and non-zero in any frame (e.g. the lab frame). However, any non-zero values are expected to arise from Planck-scale effects in the early Universe. Therefore they should be constant in a cosmological frame (e.g. the one defined by the CMB radiation) or any frame that moves with a constant velocity and shows no rotation with respect to the cosmological one. Consequently the conventionally chosen frame to analyze and compare experiments in the SME is a sun-centered, non-rotating frame as defined in \cite{KM}. The general procedure is to calculate the SME perturbation of the experimental observable in the lab frame (or cavity frame, or atom frame) and then to transform the lab frame SME parameters to the conventional sun-centered frame. This transformation will introduce a time variation of the frequency related to the movement of the lab with respect to the sun-centered frame (typically introducing time variations of sidereal and semi-sidereal periods for an Earth fixed experiment).

\subsubsection{SME photon sector}

The photon sector of the SME is described by a Lagrangian that takes the form

\begin{equation}
{\cal L} = -\frac{1}{4}F_{\mu \nu}F^{\mu \nu}+\frac{1}{2}(k_{AF})^\kappa\epsilon_{\kappa\lambda\mu\nu}A^\lambda F^{\mu\nu}
-\frac{1}{4}(k_F)_{\kappa\lambda\mu\nu}F^{\kappa \lambda}F^{\mu \nu}
\label{SMElag}
\end{equation}
where $F_{\mu\nu} \equiv \partial_\mu A_\nu - \partial_\nu A_\mu$. The first term is the usual Maxwell part while the second and third represent Lorentz violating contributions that depend on the parameters $k_{AF}$ and $k_F$. For most analysis the $k_{AF}$ parameter is set to 0 for theoretical reasons (c.f. \cite{KM}), which is also well supported experimentally. The remaining dimensionless tensor $(k_F)_{\kappa\lambda\mu\nu}$ has a total of 19 independent components that need to be determined by experiment. Retaining only this term leads to Maxwell equations that take the familiar form but with $\bf D$ and $\bf H$ fields defined by a general matrix equation

\begin{equation}
\left( \begin{array}{c}
{\bf D} \\
{\bf H} \end{array}\right)
 = \left( \begin{array}{c}
\epsilon_0(\widetilde{\epsilon_{r}}+\kappa_{DE}) \\
\sqrt{\frac{\epsilon_0}{\mu_0}}\kappa_{HE} \end{array}
\begin{array}{c}
\sqrt{\frac{\epsilon_0}{\mu_0}}\kappa_{DB} \\
\mu_0^{-1}(\widetilde{\mu_{r}}^{-1}+\kappa_{HB}) \end{array} \right)
\left( \begin{array}{c}
{\bf E} \\
{\bf B} \end{array}\right)
\label{DHdef}
\end{equation}
where the $\kappa$ are $3\times 3$ matrices whose components are particular combinations of the $k_F$ tensor (c.f. equation (5) of \cite{KM}). If we suppose the 
medium of interest has general magnetic or dielectric properties, 
then $\widetilde{\epsilon_{r}}$ and $\widetilde{\mu_{r}}$ are also 3 x 3 matrices. 
In vacuum $\widetilde{\epsilon_{r}}$ and $\widetilde{\mu_{r}}$ are identity 
matrices. Equation (\ref{DHdef}) indicates a useful analogy between the SME in vacuum and standard Maxwell equations in homogeneous anisotropic media.

For the analysis of different experiments it turns out to be useful to introduce further combinations of the $\kappa$ matrices defined by:

\begin{eqnarray}
(\tilde{\kappa}_{e+})^{jk}&=&\frac{1}{2}(\kappa_{DE}+\kappa_{HB})^{jk}, \nonumber \\
(\tilde{\kappa}_{e-})^{jk}&=&\frac{1}{2}(\kappa_{DE}-\kappa_{HB})^{jk} - \frac{1}{3}\delta^{jk}(\kappa_{DE})^{ll}, \nonumber \\
(\tilde{\kappa}_{o+})^{jk}&=&\frac{1}{2}(\kappa_{DB}+\kappa_{HE})^{jk}, \nonumber \\
(\tilde{\kappa}_{o-})^{jk}&=&\frac{1}{2}(\kappa_{DB}-\kappa_{HE})^{jk}, \nonumber \\
\tilde{\kappa}_{tr}&=&\frac{1}{3}(\kappa_{DE})^{ll}.
\label{kappadef}
\end{eqnarray}

The first four of these equations define traceless $3 \times 3$ matrices, while the last defines a single coefficient. All $\tilde{\kappa}$ matrices are symmetric except $\tilde{\kappa}_{o+}$ which is antisymmetric. These characteristics leave a total of 19 independent coefficients of the $\tilde{\kappa}$.

In general experimental results are quoted and compared using the $\tilde{\kappa}$ parameters rather than the original $k_F$ tensor components. The 10 independent components of the $\tilde{\kappa}_{e+}$ and $\tilde{\kappa}_{o-}$ tensors, have been determined to $\leq 2 \times 10^{-32}$ by astrophysical tests \cite{KM}. Of the 9 remaining independent components, 4 components of $\tilde{\kappa}_{e-}$ and the 3 components of $\tilde{\kappa}_{o+}$ have been bounded by the resonator experiments reported here and in \cite{Muller,Wolf2004} to parts in $10^{15}$ and $10^{11}$ respectively, with our results improving by up to a factor 10 on the best previous ones (c.f. Tab.\ref{Tab4}). The scalar $\tilde{\kappa}_{tr}$ has been bounded recently by our SME analysis \cite{Tobar2004} of the experiment of \cite{Saathoff} to parts in $10^{-5}$. In \cite{Tobar2004} we also propose several interferometer and resonator experiments that could improve the limit on $\tilde{\kappa}_{tr}$ to parts in $10^{11}$ and the limits on $\tilde{\kappa}_{o+}$ to parts in $10^{15}$. Finally, the remaining component $\tilde{\kappa}^{ZZ}_{e-}$ is undetermined at present as it is not accessible to Earth fixed experiments. However, it should be accessible to experiments that are rotating in the laboratory, like the ones reported elsewhere in these proceedings, which should yield the first limits on that parameter and thereby complete the coverage of the parameter space in the SME photon sector. Present limits are summarized in Tab.\ref{SMEphoton} (not including new limits reported in these proceedings). 

\begin{table}
\caption{Present limits on Lorentz violating parameters in the
SME photon sector, not including new limits reported in these proceedings.}
\begin{center}
\renewcommand{\arraystretch}{1.4}
\setlength\tabcolsep{5pt}
\begin{tabular}{ccccccc}
\hline\hline
Parameter & $\tilde{\kappa}_{e+}$ & $\tilde{\kappa}_{o-}$ &$\tilde{\kappa}_{e-}$& $(\tilde{\kappa}_{e-}^{ZZ})$ & $\tilde{\kappa}_{o+}$ & $\tilde{\kappa}_{tr}$\\
No. of components & 5 & 5 & 4 & 1 & 3 & 1 \\
Limits & $10^{-32}$ & $10^{-32}$ & $10^{-15}$ & - & $10^{-11}$ & $10^{-5}$ \\
Reference & \cite{KM}& \cite{KM} & \cite{Wolf2004,Muller} & - & \cite{Wolf2004,Muller} & \cite{Tobar2004} \\
\hline\hline
\end{tabular}
\end{center}
\label{SMEphoton}
\end{table}

\subsubsection{SME matter sector}

In the matter sector, the SME modifies the Lagrangian of a spin 1/2 fermion \cite{KostoCol,KL} via a number of parameterized Lorentz violating terms. When applied to atomic physics, this leads to a perturbation of the standard model Hamiltonian parametrised by 40 parameters for each fundamental particle (proton, neutron, electron), which in turn leads to a shift of the atomic energy levels and atomic transition frequencies (see \cite{KL,Bluhm} for details). Quite generally, the energy level shifts can be expressed in the form

\begin{equation}
\label{DE_SME}
\Delta E = \hat{m}_F(E_d^e+E_d^p+E_d^n) + \tilde{m}_F(E_q^e+E_q^p+E_q^n)
\end{equation}
where $E_d$ and $E_q$ are energies given below, the superscripts $e,p,n$ stand for electron, proton and neutron and $\hat{m}_F$ and $\tilde{m}_F$ are defined as

\begin{equation}
\label{mF_def}
\hat{m}_F := \frac{m_F}{F}, \hspace{5mm} \tilde{m}_F := \frac{3m_F^2-F(F+1)}{3F^2-F(F+1)}.
\end{equation}

In general $\Delta E$ of (\ref{DE_SME}) will be time varying as the energies $E_d^w, E_q^w$ ($w$ stands for $e,p,n$) depend on the orientation of the angular momentum of $w$ with respect to the fixed stars (best approximation to the frame in which symmetry breaking took place in the early universe). Of particular interest will be (see section \ref{FO2}) Zeeman sublevels with $m_F \neq 0$ in which case the orientation of the quantization axis (quantization magnetic field) is relevant, so one can expect variations of $\Delta E$ at sidereal and semi-sidereal frequencies due to the precession of the quantization axis with the rotation of the Earth.

The energies in (\ref{DE_SME}) are \cite{KL}

\begin{eqnarray}
E_d^w &=& \beta_w\tilde{b}_3^w + \delta_w\tilde{d}_3^w + \kappa_w\tilde{g}_d^w \nonumber \\
E_q^w &=& \gamma_w\tilde{c}_q^w + \lambda_w\tilde{g}_q^w. \label{EdEq}
\end{eqnarray}

In (\ref{EdEq}) the tilde quantities have the dimensions of energy and represent laboratory frame combinations of the SME parameters that need to be determined by experiment. They are time varying at sidereal and semi-sidereal frequencies as they are obtained by transforming the constant sun-centered-frame parameters to the laboratory frame. The other coefficients in (\ref{EdEq}) are constant and depend on the nuclear and electronic structure of the atom. Explicit expressions can be found in \cite{KL}, with their values calculated for certain atoms and states (including the $^{133}$Cs atom of interest to our experiment) in \cite{Bluhm}.

Our experiment (see section \ref{FO2}) is sensitive to $\tilde{c}_q^p$. When transforming to the sun-centered-frame this parameter is a time varying combination of 8 constant SME parameters ($\tilde{c}_Q$, $\tilde{c}_-$, $\tilde{c}_X$, $\tilde{c}_Y$, $\tilde{c}_Z$, $\tilde{c}_{TX}$, $\tilde{c}_{TY}$, $\tilde{c}_{TZ}$), which are generally used \cite{KL,Bluhm} to state and compare experimental results (see Tab. \ref{SME_tab}). In some publications \cite{Muller2005,Lane2005} the results are stated in terms of dimensionless sun-frame parameters related to the $\tilde{c}$ parameters by (c.f. \cite{Bluhm} Appendix B)

\begin{eqnarray}
\tilde{c}_Q &=& mc^2(c_{XX}+c_{YY}-2c_{ZZ}) \nonumber\\
\tilde{c}_- &=& mc^2(c_{XX}-c_{YY}) \\ \label{c_to_c}
\tilde{c}_J &=& mc^2|\epsilon_{JKL}|c_{KL} \nonumber\\
\tilde{c}_{TJ} &=& mc^2(c_{TJ}+c_{JT}) \nonumber\\
\nonumber
\end{eqnarray}
where m is the mass of the particle ($m_n$, $m_p$, or $m_e$), indices $J,K,L$ run over sun-frame spatial coordinates $X,Y,Z$ and the totally antisymmetric tensor $\epsilon_{JKL}$ is defined with $\epsilon_{XYZ}=+1$.

Existing bounds on the 40 parameters for each particle (n,p,e) come from clock comparison and magnetometer experiments using different atomic species (\cite{Bluhm} and references therein, \cite{Cane}), from resonator experiments (including our experiments described in section \ref{Molly} as analyzed recently by M\"uller) \cite{Muller2005,Wolf2004,Muller}, and from analysis of Ives-Stilwell (Doppler-shift) experiments \cite{Lane2005,Saathoff}. They are summarized in Tab. \ref{SME_tab} below. The expected results of our present experiment (see section \ref{FO2}) are given in Tab. \ref{SME_tab} in brackets. They correspond to first measurements of some parameters and an improvement by 11 and 14 orders of magnitude on others.

\begin{table}
\caption{Orders of magnitude of present limits (in GeV) on Lorentz violating parameters in the
SME matter sector and corresponding references. Expected uncertainties from the experiment reported in section \ref{FO2} are given in brackets.}
\begin{center}
\renewcommand{\arraystretch}{1.4}
\setlength\tabcolsep{5pt}
\begin{tabular}{ccccc}
\hline\hline\noalign{\smallskip}
Parameter & Proton & Neutron & Electron & References\\
\hline\noalign{\smallskip}
$\tilde{b}_X$, $\tilde{b}_Y$ & $10^{-27}$ & $10^{-31}$ & $10^{-29}$ & \cite{Phillips}, \cite{Bear}, \cite{Hou} \\
$\tilde{b}_Z$ & - & - & $10^{-28}$ & \cite{Hou} \\
$\tilde{b}_T$, $\tilde{g}_T$, $\tilde{H}_{JT}$, $\tilde{d}_\pm$, $\tilde{d}_Q$, $\tilde{d}_{XY}$, $\tilde{d}_{YZ}$ & - &  $10^{-27}$ & - & \cite{Cane}\\
$\tilde{d}_{XZ}$ & - & - & - \\
$\tilde{d}_X$, $\tilde{d}_Y$ & $10^{-25}$ & $10^{-29}$ & $10^{-22}$ & \cite{KL}, \cite{Cane}, \cite{KL}\\
$\tilde{d}_{Z}$ & - & - & - \\
$\tilde{g}_{DX}$, $\tilde{g}_{DY}$ & $10^{-25}$ & $10^{-29}$ & $10^{-22}$ &\cite{KL}, \cite{Cane}, \cite{KL}\\
$\tilde{g}_{DZ}$ & - & - & - \\
$\tilde{g}_{JK}$ & - & - & - \\
$\tilde{g}_{c}$ & - & $10^{-27}$ & - & \cite{Cane}\\
$\tilde{g}_{-}$, $\tilde{g}_{Q}$, $\tilde{g}_{TJ}$ & - & - & - \\
$\tilde{c}_{-}$ & $(10^{-25})$ & $10^{-27}$ & $10^{-19}$ & \cite{KL}, \cite{Muller2005,Wolf2004,Muller} \\
$\tilde{c}_{Q}$ & $(10^{-25})$ $10^{-11}$ & - & $10^{-9}$ & \cite{Lane2005,Saathoff} \\
$\tilde{c}_X$, $\tilde{c}_Y$ & $(10^{-25})$ & $10^{-25}$ & $10^{-19}$ &  \cite{KL}, \cite{Muller2005,Wolf2004,Muller} \\
$\tilde{c}_Z$ & $(10^{-25})$ & $10^{-27}$ & $10^{-19}$ &  \cite{KL}, \cite{Muller2005,Wolf2004,Muller} \\
$\tilde{c}_{TJ}$ & $(10^{-19})$ $10^{-8}$ & - & $10^{-6}$ & \cite{Lane2005,Saathoff} \\
\hline\hline
\end{tabular}
\end{center}
\label{SME_tab}
\end{table}

\section{Michelson-Morley and Kennedy-Thorndike tests} \label{Molly}

In this section we review the results \cite{Wolf2003,WolfGRG,Wolf2004} of our experiment that compares the frequencies of a cryogenic sapphire oscillator (CSO) and a hydrogen maser atomic clock. Both devices operate at microwave frequencies and are run and compared continuously for timekeeping purposes at the Paris observatory. We use that data to carry out Michelson-Morley and Kennedy-Thorndike experiments, searching for a dependence of the difference frequency on the orientation and/or the velocity of the CSO with respect to a prefered frame candidate.

\begin{figure}[b]
\begin{center}
\includegraphics[width=9cm]{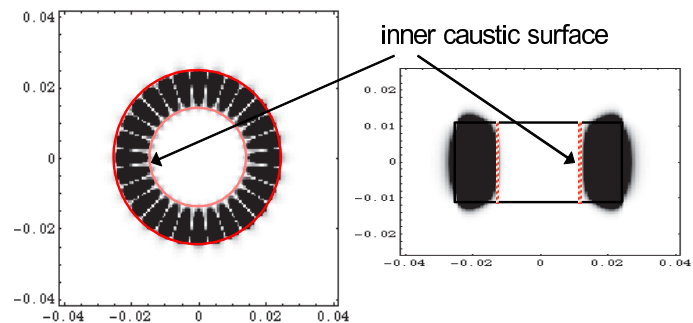}
\end{center}
\caption[]{Magnitude of the $H_z$ field calculated for the WG$_{14,0,0}$ mode in a sapphire disk resonator. The r-$\phi$ and r-z planes are represented. The inner caustic is shown, and the mode can be approximated as two guided waves propagating in opposite directions around the azimuth.}
\label{MMKTfig1}
\end{figure}

The heart of the experiment is a monolithic sapphire crystal of cylindrical shape, about 5 cm diameter and 3 cm height. The resonance frequency is determined by exciting a so called Whispering Gallery mode, corresponding to a standing wave set up around the perimeter of the cylinder (see fig. \ref{MMKTfig1} and \cite{WolfGRG} for a detailed description). In our case the excited mode is a TE mode at 11.932 GHz, with dominant radial electric and vertical magnetic fields corresponding to propagation (Poynting) vectors in both directions around the circumference. The CSO is an active system oscillating at the resonant frequency (i.e. a classical loop oscillator which amplifies and re-injects the "natural" resonator signal). Additionally the signal is locked to the resonance using the Pound-Drever technique (modulation at $\approx$ 80 kHz). The incident power is stabilized in the cryogenic environment and the spurious AM modulation is minimized using a servo loop. To minimize temperature sensitivity the resonator is heated (inside the 4 K environment) and stabilized to the temperature turning point ($\approx$ 6 K) of the resonator frequency which arises due to paramagnetic impurities in the sapphire. Under these conditions the loaded quality factor of the resonator is slightly below $10^9$. The resonator is kept permanently at cryogenic temperatures, with helium refills taking place about every 20 - 25 days.

The CSO is compared to a commercial (Datum Inc.) active hydrogen maser whose frequency is also regularly compared to caesium and rubidium atomic fountain clocks in the laboratory \cite{Bize}. The CSO resonant frequency at 11.932 GHz is compared to the 100 MHz output of the hydrogen maser. The maser signal is multiplied up to 12 GHz of which the CSO signal is subtracted. The  remaining $\approx$ 67 MHz signal is mixed to a synthesizer signal at the same frequency and the low frequency beat at $\approx$ 64 Hz is counted, giving access to the frequency difference between the maser and the CSO. The instability of the comparison chain has been measured at $\leq 2 \times 10^{-14}\tau^{-1}$, with long term instabilities dominated by temperature variations, but not exceeding $10^{-16}$.

\subsection{Results in the RMS framework} \label{MollyRMS}

In the RMS framework our experiment sets the most stringent limit for Kennedy-Thorndike experiments (improving by a factor 70 over previous results) and is among the most precise Michelson-Morley tests (see table \ref{MStab}). Those results were reported in \cite{Wolf2003,WolfGRG} and are summarized here.

In the RMS framework the frequency of a resonator in the lab frame $S$ is proportional to ${t_c}^{-1}$ where $t_c$ is the return travel time of a light signal in the resonator. Setting
$c^2dT^2=dX^2+dY^2+dZ^2$ in the preferred frame $\Sigma$, and transforming according
to (\ref{MStransf}) we find the coordinate travel time of a light
signal in $S$:

\begin{equation}
dt={dl\over c}\left(1-\left(\beta_{\mathrm{MS}}
-\alpha_{\mathrm{MS}} -1 \right){v^2\over c^2} - \left({1\over
2}-\beta_{\mathrm{MS}} +\delta_{\mathrm{MS}} \right){\rm
sin}^2\theta{v^2\over c^2}\right)+{\cal O}(4) \label{MSc}
\end{equation}
where $dl = \sqrt{dx^2+dy^2+dz^2}$ and $\theta$ is the angle
between the direction of light propagation and the velocity {\bf
v} of $S$ in $\Sigma$.

Calculating $t_c$ from (\ref{MSc}) the relative
frequency difference between the sapphire oscillator and the
hydrogen maser (which, by definition, realizes coordinate time in
$S$ \cite{masercom}) is

\begin{equation}
{\Delta\nu (t) \over \nu_0} = P_{KT}{v(t)^2\over c^2} +
P_{MM}{v(t)^2\over c^2}{1 \over 2\pi}\int_0^{2 \pi}{\rm
sin}^2\theta (t,\varphi ) d\varphi +{\cal O}(3) \label{MSdff}
\end{equation}
where $\nu_0$ is the unperturbed frequency, $v(t)$ is the (time dependent) speed
of the lab in $\Sigma$, and $\varphi$ is the azimuthal angle of
the light signal in the plane of the cylinder. The periodic time
dependence of $v$ and $\theta$ due to the rotation and orbital
motion of the Earth with respect to the CMB frame allow us to set
limits on the two parameters in (\ref{MSdff}) by fitting the
periodic terms of appropriate frequency and phase (see \cite{Mike}
for calculations of similar effects for several types of
oscillator modes). Given the limited durations of our data sets
($\leq$ 16 days) the dominant periodic terms arise from the
Earth's rotation, so retaining only those we have ${\bf v}(t) =
{\bf u}+{\bf \omega} \times {\bf R}$ with ${\bf u}$ the velocity
of the solar system with respect to the CMB, ${\bf \omega}$ the
angular velocity of the Earth, and ${\bf R}$ the geocentric
position of the lab. We then find after some calculation.

\begin{equation}
\begin{array}{cl}
\Delta\nu / \nu_0 &= P_{KT}(H{\rm sin}\lambda )\\
\ & + P_{MM}(A{\rm cos}\lambda + B{\rm cos}(2\lambda)+C{\rm
sin}\lambda+D{\rm sin}\lambda{\rm cos}\lambda+E{\rm
sin}\lambda{\rm cos}(2\lambda))
\end{array}
\label{MSdff2}
\end{equation}
where $\lambda =\omega t + \phi$, and A-E and $\phi$ are constants
depending on the latitude and longitude of the lab $(\approx 48.7
^\circ$N and $2.33 ^\circ$E for Paris). Numerically $H \approx
-2.6 \times 10^{-9}$, $A \approx -8.8 \times 10^{-8}$, $B \approx
1.8 \times 10^{-7}$, C-E of order $10^{-9}$. We note that in
(\ref{MSdff2}) the dominant time variations of the two
combinations of parameters are in quadrature and at twice the
frequency which indicates that they should decorelate well in the
data analysis allowing a simultaneous determination of the two (as
confirmed by the correlation coefficients given below).
Fitting this simplified model to our data we obtain results that
differ by less than 10\% from the results presented below
that were obtained using the complete model ((\ref{MSdff}) including the orbital motion of the Earth).
 
For the RMS analysis we use 13 data sets in total spanning Sept. 2002 to Aug. 2003, of differing lengths (5 to 16 days, 140 days in total). The sampling time for all data sets was $100$ s except two data sets with $\tau_0 = 12$ s. To make the data more manageable we first average all points to $\tau_0 = 2500$ s. For the data analysis we simultaneously fit (using weighted least squares, WLS, c.f. \cite{WolfGRG}) an offset and a rate (natural frequency drift, typically $\approx 1.7 \times 10^{-18}$ s$^{-1}$) per data set and the two parameters of the model (\ref{MSdff}). In the model (\ref{MSdff}) we take into account the rotation of the Earth and the Earth's orbital motion, the latter contributing little as any constant or linear terms over the durations of the individual data sets are absorbed by the fitted offsets and rates.

\begin{figure}[b]
\begin{center}
\includegraphics[width=9cm]{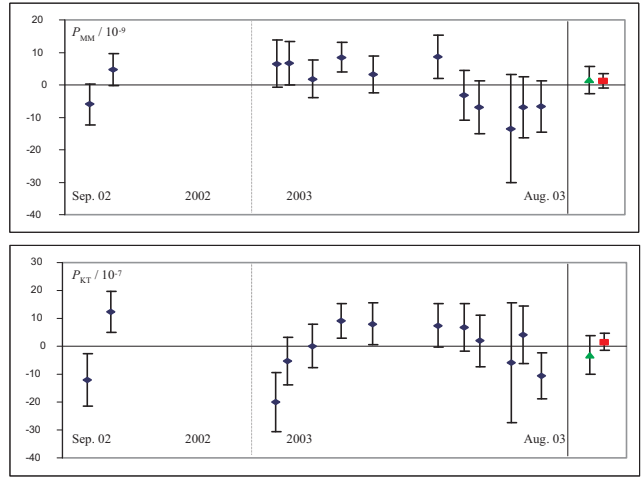}
\end{center}
\caption[]{Values (published in \cite{WolfGRG}) of the two parameters ($P_{KT}$ and $P_{MM}$) from a fit to each individual data set (blue diamonds) and a global fit to all the data (red squares). For comparison the previous results published in \cite{Wolf2003} are also shown (green triangles). The error bars indicate the combined uncertainties from statistics and systematic effects.}
\label{MMKTfig2}
\end{figure}

Figure \ref{MMKTfig2} shows the resulting values of the two parameters ($P_{KT}$ and $P_{MM}$) for each individual data set. A global WLS fit of the two parameters and the 13 offsets and drifts yields $P_{MM} = (1.2\pm 1.9) \times 10^{-9}$ and $P_{KT} = (1.6\pm 2.3) \times 10^{-7}$ ($1\sigma$ uncertainties), with the correlation coefficient between the two parameters less than 0.01 and all other correlation coefficients $< 0.06$. The distribution of the 13 individual values around the ones obtained from the global fit is well compatible with a normal distribution ($\chi^2$ = 10.7 and $\chi^2$ = 14.6 for $P_{MM}$ and $P_{KT}$ respectively).

Systematic effects at diurnal or semi-diurnal frequencies with the appropriate phase could mask a putative sidereal signal. The statistical uncertainties of $P_{MM}$ and $P_{KT}$ obtained from the WLS fit above correspond to sidereal and semi-sidereal terms (from (\ref{MSdff2})) of $\approx 7 \times 10^{-16}$ and $\approx 4 \times 10^{-16}$ respectively so any systematic effects exceeding these limits need to be taken into account in the final uncertainty. We expect the main contributions to such effects to arise from temperature, pressure and magnetic field variations that would affect the hydrogen maser, the CSO and the associated electronics, and from tilt variations of the CSO which are known to affect its frequency (see section \ref{MollySME} for a detailed discussion). Our final uncertainties (the error bars in Fig. \ref{MMKTfig2}) are the quadratic sums of the statistical uncertainties from the WLS adjustment for each data set and the systematic uncertainties calculated for each data set from (\ref{MSdff2}). For the global adjustment we average the systematic uncertainties from the individual data sets obtaining $\pm 1.2 \times 10^{-9}$ on $P_{MM}$ and $\pm 1.9 \times 10^{-7}$ on $P_{KT}$.

In the RMS framework, our experiment simultaneously constrains two combinations of the three parameters of the Mansouri and Sexl test theory (previously measured individually by Michelson-Morley and Kennedy-Thorndike experiments). We obtain $\delta_{\mathrm{MS}} - \beta_{\mathrm{MS}} + 1/2 = 1.2(1.9)(1.2) \times 10^{-9}$ which is of the same order as the best previous results \cite{Muller,Brillet}, and $\beta_{\mathrm{MS}} - \alpha_{\mathrm{MS}} - 1 = 1.6(2.3)(1.9)\times 10^{-7}$ which improves the best previous limit \cite{Schiller} by a factor of 70 (the first bracket indicates the $1\sigma$ uncertainty from statistics the second from systematic effects). We note that our value on $\delta_{\mathrm{MS}} - \beta_{\mathrm{MS}} + 1/2$ is compatible with the slightly significant recent result of \cite{Muller} who obtained $\delta_{\mathrm{MS}} - \beta_{\mathrm{MS}} + 1/2 = (2.2 \pm 1.5)\times 10^{-9}$. 

As a result of our experiment the Lorentz transformations are confirmed in the RMS framework (c.f. Tab. \ref{MStab}) with an overall uncertainty of $\leq 3 \times 10^{-7}$ limited by our determination of $\beta_{\mathrm{MS}} - \alpha_{\mathrm{MS}} - 1$ and the recent limit \cite{Saathoff} of $2.2 \times 10^{-7}$ on the determination of $\alpha_{\mathrm{MS}}$. The latter is likely to improve in the coming years by experiments such as ACES (Atomic
Clock Ensemble in Space \cite{ACES}) that will compare ground
clocks to clocks on the international space station aiming at a $10^{-8}$ measurement of $\alpha_{\mathrm{MS}}$.

\subsection{Results in the SME} \label{MollySME}

In the SME our experiment sets the presently most stringent limits on a number of photon sector parameters, improving previous results \cite{Muller} by up to an order of magnitude. These results were first published in \cite{Wolf2004} and are reproduced here.

The SME perturbed frequency of a resonator can be calculated from equation (\ref{DHdef}) in the form (c.f. \cite{KM})

\begin{eqnarray}
\frac{\Delta\nu}{\nu_0} = &-&\frac{1}{\langle U \rangle} \int_V d^3x \left(\epsilon_0 {\bf E_0}^*\cdot \kappa_{DE}\cdot {\bf E_0} - \mu_0^{-1} {\bf B_0}^*\cdot \kappa_{HB}\cdot {\bf B_0} \right. \\
&+& \left. 2{\rm Re}(\sqrt{\frac{\epsilon_0}{\mu_0}}{\bf E_0}^*\cdot \kappa_{DB}\cdot {\bf B_0}) \right) \nonumber
\label{KM34}
\end{eqnarray}
where ${\bf B_0}, {\bf H_0}, {\bf E_0}, {\bf D_0}$ are the unperturbed (standard Maxwell) fields and $\langle U \rangle = \int_V d^3x ({\bf E_0}\cdot {\bf D_0}^*+{\bf B_0}\cdot {\bf H_0}^*)$. Note that, as shown in \cite{KL}, the frequency of the H-maser is not affected to first order (because it operates on $m_F=0$ states) and \cite{Muller2} shows that the perturbation of the frequency due to the modification of the sapphire crystal structure (and hence the cavity size) is negligible with respect to the direct perturbation of the e-m fields.

The resonator is placed in the lab with its symmetry axis along the vertical. Applying (\ref{KM34}) in the lab frame (z-axis vertical upwards, x-axis pointing south), with the fields calculated using a finite element technique as described in \cite{WolfGRG}, we obtain an expression for the frequency variation of the resonator

\begin{eqnarray}
\frac{\Delta\nu}{\nu_0}&=& ({\cal M}_{DE})_{lab}^{xx}\left((\kappa_{DE})_{lab}^{xx}+(\kappa_{DE})_{lab}^{yy}\right)+({\cal M}_{DE})_{lab}^{zz}(\kappa_{DE})_{lab}^{zz} \nonumber \\
&+& ({\cal M}_{HB})_{lab}^{xx}\left((\kappa_{HB})_{lab}^{xx}+(\kappa_{HB})_{lab}^{yy}\right)+({\cal M}_{HB})_{lab}^{zz}(\kappa_{HB})_{lab}^{zz}
\label{dffM}
\end{eqnarray}
with the ${\cal M}_{lab}$ components given in Tab. \ref{tab:Mike5}. To obtain the values in Tab. \ref{tab:Mike5} we take into account the fields inside the resonator (c.f. \cite{WolfGRG}) and outside ($\leq 2\%$ of the energy).

\begin{table}[htb]
\caption{${\cal M}_{lab}$ components calculated using (\ref{KM34}) and a finite element technique for the determination of the fields inside the resonator (see \cite{WolfGRG} for details)}
\begin{center}
\renewcommand{\arraystretch}{1.4}
\setlength\tabcolsep{5pt}
\begin{tabular}{cccc}
\hline \hline
$({\cal M}_{DE})_{lab}^{xx}$ & $({\cal M}_{DE})_{lab}^{zz}$ & $({\cal M}_{HB})_{lab}^{xx}$ & $({\cal M}_{HB})_{lab}^{zz}$ \\
\hline
-0.03093 \ & \ -0.0004030\ & \ 0.008408\ & \ 0.4832\ \\
\hline \hline
\end{tabular}
\end{center}
\label{tab:Mike5}
\end{table}

The last step is to transform the $\kappa$ tensors in (\ref{dffM}) to the conventional sun-centered frame using the explicit transformations provided in \cite{KM}, and to express the result in terms of the $\tilde{\kappa}$ tensors of (\ref{kappadef}). We obtain

\begin{equation}
\frac{\nu-\nu_0}{\nu_0} = \sum_i C_i {\rm cos}(\omega_i T_\oplus + \varphi_i) + S_i {\rm sin}(\omega_i T_\oplus + \varphi_i)\label{nuSME}
\end{equation}
where $\nu_0$ is the unperturbed frequency difference, the sum is over the six frequencies $\omega_i$ of Tab.\ref{Tab3}, the coefficients $C_i$ and $S_i$ are functions of the Lorentz violating tensors $\tilde{\kappa}_{e-}$ and $\tilde{\kappa}_{o+}$ (see Tab.\ref{Tab3}), $T_\oplus = 0$ on December 17, 2001, 18:05:16 UTC, $\varphi_{\omega_\oplus}=\varphi_{2\omega_\oplus}=0$ and $\varphi_{(\omega_\oplus\pm\Omega_\oplus)}=\varphi_{(2\omega_\oplus\pm\Omega_\oplus)}=\pm 4.682$ rad. To obtain the relations of Tab.\ref{Tab3} between $C_i$, $S_i$ and the SME parameters we have assumed zero values for the 10 independent components of the $\tilde{\kappa}_{e+}$ and $\tilde{\kappa}_{o-}$ tensors, as those have been determined to $\leq 2 \times 10^{-32}$ by astrophysical tests \cite{KM}.

\begin{table*}
\caption{\label{Tab3} Coefficients $C_i$ and $S_i$ in (1) for the six frequencies $\omega_i$ of interest and their relation to the components of the SME parameters $\tilde{\kappa}_{e-}$ and $\tilde{\kappa}_{o+}$, with $\omega_\oplus$ and $\Omega_\oplus$ the angular frequencies of the Earth's sidereal rotation and orbital motion. The measured values (in $10^{-16}$) are shown together with the statistical (first bracket) and systematic (second bracket) uncertainties.}
\begin{center}
\renewcommand{\arraystretch}{1.4}
\setlength\tabcolsep{5pt}
\begin{tabular}{ccc}
\hline \hline
$\omega_i$ &$C_i$&$S_i$\\
\hline
$\omega_\oplus - \Omega_\oplus$ & $(-8.6\times 10^{-6})\tilde{\kappa}_{o+}^{YZ}$ & $(8.6\times 10^{-6})\tilde{\kappa}_{o+}^{XZ}-(4.2\times 10^{-5})\tilde{\kappa}_{o+}^{XY}$ \\
$\omega_\oplus$ & $-0.44\tilde{\kappa}_{e-}^{XZ}+(1.1\times 10^{-6})\tilde{\kappa}_{o+}^{XZ}$ & $-0.44\tilde{\kappa}_{e-}^{YZ}+(1.1\times 10^{-6})\tilde{\kappa}_{o+}^{YZ}$ \\
$\omega_\oplus + \Omega_\oplus$ & $(-8.6\times 10^{-6})\tilde{\kappa}_{o+}^{YZ}$ & $(8.6\times 10^{-6})\tilde{\kappa}_{o+}^{XZ}+(1.8\times 10^{-6})\tilde{\kappa}_{o+}^{XY}$ \\
$2\omega_\oplus - \Omega_\oplus$ & $(-1.8\times 10^{-5})\tilde{\kappa}_{o+}^{XZ}$ & $(-1.8\times 10^{-5})\tilde{\kappa}_{o+}^{YZ}$ \\
$2\omega_\oplus$ & $-0.10(\tilde{\kappa}_{e-}^{XX}-\tilde{\kappa}_{e-}^{YY})$ & $-0.19\tilde{\kappa}_{e-}^{XY}$ \\
$2\omega_\oplus + \Omega_\oplus$ & $(7.8\times 10^{-7})\tilde{\kappa}_{o+}^{XZ}$ & $(7.8\times 10^{-7})\tilde{\kappa}_{o+}^{YZ}$ \\
\hline\hline
$\omega_\oplus - \Omega_\oplus$ & $-6.9(4.2)(4.5)$ & $6.7(4.2)(4.5)$\\
$\omega_\oplus$ & $14(4.2)(4.2)$ & $2.4(4.2)(4.2)$\\
$\omega_\oplus + \Omega_\oplus$ & $-6.0(4.2)(4.2)$ & $2.7(4.2)(4.2)$\\
$2\omega_\oplus - \Omega_\oplus$ & $3.7(2.4)(3.7)$ & $-2.9(2.4)(3.7)$\\
$2\omega_\oplus$ & $3.1(2.4)(3.7)$ & $11(2.4)(3.7)$\\
$2\omega_\oplus + \Omega_\oplus$ & $0.0(2.4)(3.7)$ & $-1.2(2.4)(3.7)$\\
\hline\hline
\end{tabular}
\end{center}
\end{table*}

To determine all 7 SME parameters appearing in Tab.\ref{Tab3} one requires over a year of data in order to be able to decorrelate the annual sidebands from the sidereal and twice sidereal frequencies. To do so we have extended the data to 20 data sets in total, spanning Sept. 2002 to Jan. 2004, of differing lengths (5 to 20 days, 222 days in total). The sampling time for all data sets was $100$ s.

For the statistical analysis we first average the data to 2500 s sampling time and then simultaneously fit the 20 rates and offsets and the 12 parameters $C_i$ and $S_i$ of (\ref{nuSME}) to the complete data using two statistical methods, weighted least squares (WLS), which allows one to account for non-white noise processes (cf. \cite{Wolf2003}), and individual periods (IP) as used in \cite{Muller}. The two methods give similar results for the parameters (within the uncertainties) but differ in the estimated uncertainties (the IP uncertainties are a factor $\approx 1.2$ larger). Because IP discards a significant amount of data (about 10\% in our case) we consider WLS the more realistic method and retain those results as the statistical uncertainties shown in Tab.\ref{Tab3}. We note that we now have sufficient data to decorrelate all 12 parameters ($C_i$, $S_i$) i.e. the WLS correlation coefficients between any two parameters or between any parameter and the fitted offsets and rates are all less than 0.20.

\begin{figure}[b]
\begin{center}
\includegraphics[width=10cm]{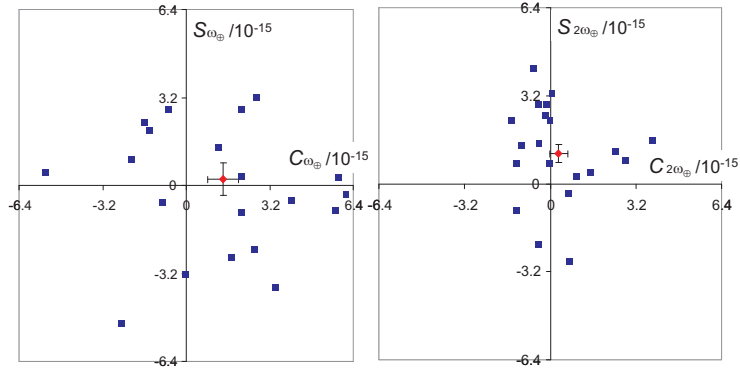}
\end{center}
\caption[]{Fitted sine and cosine amplitudes at $\omega_\oplus$ and $2\omega_\oplus$ for each data set (blue squares) and the complete data (red diamonds, with statistical errors). For clarity the error bars of the individual data sets have been omitted.}
\label{SMEfig2}
\end{figure}

\begin{figure}[b]
\begin{center}
\includegraphics[width=10cm]{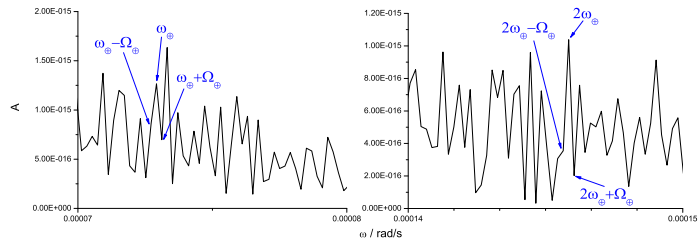}
\end{center}
\caption[]{Fitted Amplitudes $A_\omega$ for a range of frequencies around the six frequencies $\omega_i$ of interest (indicated by arrows).}
\label{SMEfig3}
\end{figure}

To investigate the distributions of our results we fit the coefficients $C_i$ and $S_i$ to each one of the 20 data sets individually with the results at the sidereal and semi-sidereal frequencies $\omega_\oplus$ and $2\omega_\oplus$ shown in Fig.\ref{SMEfig2}. If a genuine effect at those frequencies was present we would expect correlated phases of the individual points in Fig.\ref{SMEfig2}, but this does not seem to be supported by the data. A distribution of the phases may result from an effect at a neighboring frequency, in particular the diurnal and semi-diurnal frequencies $\omega_\oplus - \Omega_\oplus$ and $2(\omega_\oplus - \Omega_\oplus)$ at which we would expect systematic effects to play an important role. Fig. \ref{SMEfig3} shows the amplitudes $A_\omega = \sqrt{C_\omega^2+S_\omega^2}$ resulting from least squares fits for a range of frequencies, $\omega$, around the frequencies of interest. We note that the fitted amplitudes at $\omega_\oplus - \Omega_\oplus$ and $2(\omega_\oplus - \Omega_\oplus)$ are substantially smaller than those at $\omega_\oplus$ and $2\omega_\oplus$ and therefore unlikely to contribute to the distribution of the points in Fig.\ref{SMEfig2}.

Systematic effects at the frequencies $\omega_i$ could mask a putative Lorentz violating signal in our experiment and need to be investigated in order to be able to confirm such a signal or to exclude it within realistic limits. We have extensively studied all systematic effects arising from environmental factors that might affect our experiment. The resulting estimated contributions at the two central frequencies $\omega_\oplus$, $2\omega_\oplus$ and at the diurnal frequency $\omega_\oplus - \Omega_\oplus$ are summarized in Tab.\ref{Tab2}. The contributions at $\omega_\oplus + \Omega_\oplus$ and $2\omega_\oplus \pm \Omega_\oplus$ are not shown as they are identical to those at $\omega_\oplus$ and $2\omega_\oplus$ respectively.

\begin{table}
\caption{\label{Tab2}  Contributions from systematic effects to the amplitudes $A_i$ (parts in $10^{16}$) at three frequencies $\omega_i$.}
\begin{center}
\renewcommand{\arraystretch}{1.4}
\setlength\tabcolsep{5pt}
\begin{tabular}{lccc}
\hline\hline
Effect & $\omega_\oplus - \Omega_\oplus$ & $\omega_\oplus$ & $2\omega_\oplus$ \\
\hline
H-maser & $< 5$ & $< 5$ & $< 5$\\
Tilt & 3 & 3 & 1\\
Gravity & 0.3 & 0.3 & 0.3\\
B-field & $< 0.1$ & $< 0.1$ & $< 0.1$\\
Temperature & $< 1$ & $< 1$ & $< 1$\\
Atm. Pressure & 2.3 & 0.3 & 0.4\\
\hline
{\bf Total} & {\bf 6.4} & {\bf 5.9} & {\bf 5.2}\\
\hline\hline
\end{tabular}
\end{center}
\end{table}

We have compared the Hydrogen-maser (HM) used as our frequency reference to our highly stable and accurate Cs fountain clocks (FO2 and FOM). For example, the amplitudes at $\omega_\oplus$ and $2\omega_\oplus$ of the HM-FOM relative frequency difference over June-July 2003 were $A_{\omega_\oplus}=(4.8\pm 4.7)\times 10^{-16}$ and $A_{2\omega_\oplus}=(4.3\pm 4.7)\times 10^{-16}$. This indicates that any environmental effects on the HM at those frequencies should be below 5 parts in $10^{16}$ in amplitude. This is in good agreement with studies on similar HMs carried out in \cite{Parker} that limited environmental effects to $<$ 3 to 4 parts in $10^{16}$.

To estimate the tilt sensitivity we have intentionally tilted the oscillator by $\approx$ 5 mrad off its average position which led to relative frequency variations of $\approx 3 \times 10^{-13}$ from which we deduce a tilt sensitivity of $\approx 6 \times 10^{-17} \mu$rad$^{-1}$. This is in good agreement with similar measurements in \cite{ChangTh} that obtained sensitivities of $\approx 4 \times 10^{-17} \mu$rad$^{-1}$. Measured tilt variations in the lab at diurnal and semi-diurnal periods show amplitudes of 4.6 $\mu$rad and 1.6 $\mu$rad respectively which leads to frequency variations that do not exceed $3 \times 10^{-16}$ and $1 \times 10^{-16}$ respectively.

From the measurements of tilt sensitivity one can deduce the sensitivity to gravity variations (cf. \cite{ChangTh}), which in our case lead to a sensitivity of $\approx 3 \times 10^{-10} g^{-1}$. Tidal gravity variations can reach $\approx 10^{-7} g$ from which we obtain a maximum effect of $3 \times 10^{-17}$, one order of magnitude below the effect from tilt variations.

Variations of the ambient magnetic field in our lab. are dominated by the passage of the Paris Metro, showing a strong periodicity ("quiet" periods from 1 am to 5 am). The corresponding diurnal and semi-diurnal amplitudes are $1.7 \times 10^{-4}$ G and $3.4 \times 10^{-4}$ G respectively for the vertical field component and about 10 times less for the horizontal one. To determine the magnetic sensitivity of the CSO we have applied a sinusoidal vertical field of 0.1 G amplitude with a 200 s period. Comparing the CSO frequency to the FO2 Cs-fountain we see a clear sinusoidal signal (S/N $> 2$) at the same period with an amplitude of $7.2 \times 10^{-16}$, which leads to a sensitivity of $\approx 7 \times 10^{-15}$ G$^{-1}$. Assuming a linear dependence (there is no magnetic shielding that could lead to non-linear effects) we obtain effects of only a few parts in $10^{-18}$.

Late 2002 we implemented an active temperature stabilization inside an isolated volume ($\approx 15 {\rm m}^3$) that includes the CSO and all the associated electronics. The temperature is measured continously in two fixed locations (behind the electronics rack and on top of the dewar). For the best data sets the measured temperature variations do not exceed 0.02/0.01 K in amplitude for the diurnal and semi-diurnal components. A least squares fit to all our temperature data (taken simultaneously with our frequency measurements) yields amplitudes of $A_{\omega_\oplus}=0.020$ K and $A_{2\omega_\oplus}=0.018$ K with similar values at the other frequencies $\omega_i$ of interest, including the diurnal one ($A_{\omega_\oplus - \Omega_\oplus}=0.022$ K). Inducing a strong sinusoidal temperature variation ($\approx 0.5$ K amplitude at 12 h period) leads to no clearly visible effect on the CSO frequency. Taking the noise level around the 12 h period as the maximum effect we obtain a sensitivity of $< 4 \times 10^{-15}$ per K. Using this estimate we obtain effects of $< 1 \times 10^{-16}$ at all frequencies $\omega_i$.

Finally we have investigated the sensitivity of the CSO to atmospheric pressure variations. To do so we control the pressure inside the dewar using a variable valve mounted on the He-gas exhaust. During normal operation the valve is open and the CSO operates at ambient atmospheric pressure. For the sensitivity determination we have induced a sinusoidal pressure variation ($\approx 14$ mbar amplitude at 12 h period), which resulted in a clearly visible effect on the CSO frequency corresponding to a sensitivity of $\approx 6.5 \times 10^{-16}$ mbar$^{-1}$. We have checked that the sensitivity is not significantly affected when changing the amplitude of the induced pressure variation by a factor 3. A least squares fit to atmospheric pressure data (taken simultaneously with our frequency measurements) yields amplitudes of $A_{\omega_\oplus}=0.045$ mbar and $A_{2\omega_\oplus}=0.054$ mbar with similar values at the other frequencies $\omega_i$ of interest, except the diurnal one for which $A_{\omega_\oplus - \Omega_\oplus}=0.36$ mbar. The resulting effects on the CSO frequency are given in Tab.\ref{Tab2}.

Our final results for the 7 components of $\tilde{\kappa}_{e-}$ and $\tilde{\kappa}_{o+}$ are obtained from a least squares fit to the 12 measured coefficients of Tab.\ref{Tab3}. They are summarized and compared to the results of \cite{Muller} in Tab.\ref{Tab4}.

\begin{table}
\caption{\label{Tab4}Results for the components of the SME Lorentz violation parameters $\tilde{\kappa}_{e-}$ (in $10^{-15}$) and $\tilde{\kappa}_{o+}$ (in $10^{-11}$).}
\begin{center}
\renewcommand{\arraystretch}{1.4}
\setlength\tabcolsep{5pt}
\begin{tabular}{ccccc}
\hline\hline
& $\tilde{\kappa}_{e-}^{XY}$ & $\tilde{\kappa}_{e-}^{XZ}$ & $\tilde{\kappa}_{e-}^{YZ}$ & $(\tilde{\kappa}_{e-}^{XX}-\tilde{\kappa}_{e-}^{YY})$ \\
\hline
from \cite{Muller} & 1.7(2.6) & -6.3(12.4) & 3.6(9.0) & 8.9(4.9)\\
this work & -5.7(2.3) & -3.2(1.3) & -0.5(1.3) & -3.2(4.6)\\
\hline \hline
& $\tilde{\kappa}_{o+}^{XY}$ & $\tilde{\kappa}_{o+}^{XZ}$ & $\tilde{\kappa}_{o+}^{YZ}$ &\\
\hline
from \cite{Muller} & 14(14) & -1.2(2.6) & 0.1(2.7) &\\
this work & -1.8(1.5) & -1.4(2.3) & 2.7(2.2) &\\
\hline\hline
\end{tabular}
\end{center}
\end{table}

We note that our results for $\tilde{\kappa}_{e-}^{XY}$ and $\tilde{\kappa}_{e-}^{XZ}$ are significant at about $2\sigma$, while those of \cite{Muller} are significant at about the same level for $(\tilde{\kappa}_{e-}^{XX}-\tilde{\kappa}_{e-}^{YY})$. The two experiments give compatible results for $\tilde{\kappa}_{e-}^{XZ}$ (within the $1\sigma$ uncertainties) but not for the other two parameters, so the measured values of those are unlikely to come from a common source. Another indication for a non-genuine effect comes from figures \ref{SMEfig2} and \ref{SMEfig3}, as we would expect any genuine effect to show an approximately coherent phase for the individual data sets in figure \ref{SMEfig2} and to display more prominent peaks in figure \ref{SMEfig3}.

In conclusion, we have not seen any Lorentz violating effects in the general framework of the SME, and set limits on 7 parameters of the SME photon sector (cf. Tab. \ref{Tab4}) which are up to an order of magnitude more stringent than those obtained from previous experiments \cite{Muller}. Two of the parameters are significant (at $\approx 2\sigma$). We believe that this is most likely a statistical coincidence or a neglected systematic effect. To verify this, our experiment is continuing and new, more precise experiments are under way \cite{Mike}.

\section{Atomic clock test of Lorentz invariance in the SME matter sector}\label{FO2}

For this experiment we use one of the laser cooled fountain clocks operated at the Paris observatory, the $^{133}$Cs and $^{87}$Rb double fountain FO2 \cite{BizeJPB}. We run it in Cs mode on the $|F=4\rangle \leftrightarrow |F=3\rangle$ hyperfine transition of the $6S_{1/2}$ ground state. Both hyperfine states are degenerate, with Zeeman substates $m_F=[-4,4]$ and $m_F=[-3,3]$ respectively. The clock transition used in routine operation is $|F=4,m_F=0\rangle\leftrightarrow |F=3,m_F=0\rangle$ at 9.2 GHz, which is magnetic field independent to first order. The first order magnetic field dependent Zeeman transitions ($|F=4,m_F=i\rangle\leftrightarrow |F=3,m_F=i\rangle$ with $i=\pm 1,\pm 2,\pm 3$) are used regularly for measurement and characterization of the magnetic field, necessary to correct the second order Zeeman effect of the clock transition. In routine operation the clock transition frequency stability of FO2 is $1.6\times 10^{-14}\tau^{-1/2}$, and its accuracy $7\times 10^{-16}$ \cite{BizeJPB,Marion}, the best performance of any clock at present.

In the presence of Lorentz violation the SME frequency shift of a $Cs$ $|F=4,m_F\rangle\leftrightarrow |F=3,m_F\rangle$ transition, arising from the energy level shifts described in section \ref{SMEtheo}, has been calculated explicitly in \cite{Bluhm}. It can be written in the form

\begin{eqnarray}
\label{SMEclockshift}
\hbar (\delta\omega_{SME}) &=& s_1^p\left(\beta_p\tilde{b}_3^p - \delta_p\tilde{d}_3^p + \kappa_p\tilde{g}_d^p\right)+s_2^p\left(\gamma_p\tilde{c}_q^p - \lambda_p\tilde{g}_q^p\right) \nonumber\\
&+& s_1^e\left(\beta_e\tilde{b}_3^e - \delta_e\tilde{d}_3^e + \kappa_e\tilde{g}_d^e\right)
\end{eqnarray}
where the tilde quantities are the SME matter sector parameters described in section \ref{SMEtheo}. The quantities $\beta_w, \delta_w, \kappa_w, \gamma_w, \lambda_w$ depend on the nuclear and electronic structure, and are given in table II of \cite{Bluhm}. The $s$ coefficients result from the application of the Wigner-Eckhart theorem and are also given in \cite{Bluhm}. All coefficients entering equation (\ref{SMEclockshift}) are summarized in table \ref{CScoffs}.

\begin{table}
\caption{\label{CScoffs}Coefficients entering equation (\ref{SMEclockshift}) for a $^{133}Cs$ $|F=4,m_F\rangle\leftrightarrow |F=3,m_F\rangle$ transition. $K_p = \langle p^2\rangle/m_p^2$ for the Schmidt proton and $K_e = \langle p^2\rangle/m_e^2$ for the valence electron, with $K_p\approx 10^{-2}$ and $K_e\approx 10^{-5}$ \cite{Bluhm}.}
\begin{center}
\renewcommand{\arraystretch}{1.4}
\setlength\tabcolsep{5pt}
\begin{tabular}{ccccccccccc}
\hline\hline
$\beta_p$&$\delta_p$&$\kappa_p$&$\gamma_p$&$\lambda_p$&$\beta_e$&$\delta_e$&$\kappa_e$&$s_1^p$&$s_2^p$&$s_1^e$ \\
\hline
$\frac{7}{9}$&$-\frac{7}{33}K_p$&$\frac{28}{99}K_p$&$-\frac{1}{9}K_p$&$0$&$-1$&$\frac{1}{3}K_e$&$-\frac{1}{3}K_e$&$-\frac{1}{14}m_F$&$-\frac{1}{14}m_F^2$&$\frac{1}{2}m_F$\\
\hline\hline
\end{tabular}
\end{center}
\end{table}

From equation (\ref{SMEclockshift}) and table \ref{CScoffs} we notice that all $m_F \neq 0$ Zeeman transitions are sensitive to a violation of Lorentz symmetry, but not the $m_F=0$ clock transition. So in principle a direct measurement of one of the Zeeman transitions with respect to the clock transition (used as the reference) can yield a test of Lorentz invariance. The sensitive axis of the experiment is defined by the direction of the quantization magnetic field used to separate the Zeeman substates (vertical in the case of FO2), hence the rotation of the earth provides a modulation of the Lorentz violating signal at sidereal and semi-sidereal frequencies, which could be searched for in the data.

However, in such a direct measurement the first order Zeeman shift of the $m_F \neq 0$ transition would be the dominant error source and largely degrade the sensitivity of the experiment. The complete frequency shift of a Cs hyperfine Zeeman transition is \cite{VanAud}

\begin{equation}
\label{totalshift}
\delta\omega = \delta\omega_{SME}+m_F K_Z^{(1)}B+\left(1-\frac{m_F^2}{16}\right)K_Z^{(2)}B^2+\Delta
\end{equation}
where $\delta\omega_{SME}$ is the SME frequency shift given by (\ref{SMEclockshift}), $B$ is the magnetic field seen by the atom, $K_Z^{(1)} = 44.035$~rad~s$^{-1}$~nT$^{-1}$ is the first order Zeeman coefficient, $K_Z^{(2)}=2685.75$ rad~s$^{-1}$~T$^{-2}$ is the second order coefficient, and $\Delta$ is the shift due to other systematic effects. In (\ref{totalshift}) the diurnal and semi-diurnal variations of $B$ would mimic a putative Lorentz violating signal appearing in the sidereal and semi-sidereal variations of $\delta\omega_{SME}$ and render such a measurement very uncertain.

A somewhat cleverer strategy is to take advantage of the linear dependence on $m_F$ of the first order Zeeman shift but quadratic dependence on $m_F$ of one of the SME terms (the $s_2^p$ term in (\ref{SMEclockshift})). That implies that when measuring "simultaneously" the $m_F=3$, $m_F=-3$, and $m_F=0$ transitions and forming the observable $(\omega_{+3}+\omega_{-3}-2\omega_{0})$ one should obtain a quantity that is independent of the first order Zeeman shift, but still shows a deviation from zero and a sidereal and semi-sidereal modulation in the presence of Lorentz violation. Using (\ref{SMEclockshift}) and (\ref{totalshift}) this observable is

\begin{equation}
\label{obsshift}
(\omega_{+3}+\omega_{-3}-2\omega_{0})= \frac{1}{7}K_p\tilde{c}_q^p+K_{Z(obs)}^{(2)}B^2+\Delta_{(obs)}
\end{equation}
where $K_{Z(obs)}^{(2)}$ and $\Delta_{obs}$ are now the second order Zeeman coefficient and correction from other systematic effects for the complete observable.

The first term of (\ref{obsshift}) characterizes a possible Lorentz violation in the SME and is time varying when transforming the lab frame parameter $\tilde{c}_q^p$ to the conventional sun-centered frame. The general form of that transformation yields \cite{Bluhm}

\begin{equation}
\label{c_transform}
\tilde{c}_q^p=\tilde{B}+\tilde{C}_{\omega_\oplus}{\rm cos}(\omega_\oplus t)+\tilde{S}_{\omega_\oplus}{\rm sin}(\omega_\oplus t)+\tilde{C}_{2\omega_\oplus}{\rm cos}(2\omega_\oplus t)+\tilde{S}_{2\omega_\oplus}{\rm sin}(2\omega_\oplus t)
\end{equation}
where $\omega_\oplus$ is the frequency of rotation of the Earth. The coefficients $\tilde{B}$, $\tilde{C}_{\omega_\oplus}$, $\tilde{S}_{\omega_\oplus}$, $\tilde{C}_{2\omega_\oplus}$, $\tilde{S}_{2\omega_\oplus}$ are functions of the 8 constant sun frame SME parameters $\tilde{c}_X^p$, $\tilde{c}_Y^p$, $\tilde{c}_Z^p$, $\tilde{c}_Q^p$, $\tilde{c}_-^p$, $\tilde{c}_{TX}^p$, $\tilde{c}_{TY}^p$, $\tilde{c}_{TZ}^p$ (see \cite{Bluhm} for details) with the three $\tilde{c}_{TJ}^p$ components suppressed by a factor $v_R/c \approx 10^{-6}$ related to the velocity $v_R$ of the lab due to the rotation of the Earth.

The observable we use (equation (\ref{obsshift})) should be independent of any long term ($>$ few seconds) variations of the first order Zeeman effect and therefore any sidereal or semi-sidereal variation of the observable would be the result of Lorentz violation, if it exceeds the measurement noise and the limits imposed by other systematic effects (see below).
 
The FO2 setup is sketched in Fig.\ref{fig:fountain}. Cs atoms
effusing from an oven are slowed using a counter propagating laser
beam and captured in a lin $\perp$ lin optical molasses. Atoms are
cooled by six laser beams supplied by preadjusted fiber couplers
precisely attached to the vacuum tank and aligned along the axes of
a 3 dimensional coordinate system, where the (111) direction is
vertical. Compared to typical clock operation \cite{BizeJPB}, the number of
atoms loaded in the optical molasses has been reduced to $2\times
10^{7}$ atoms captured in 30~ms.

\begin{figure}[b]
\begin{center}
\includegraphics[width=7cm]{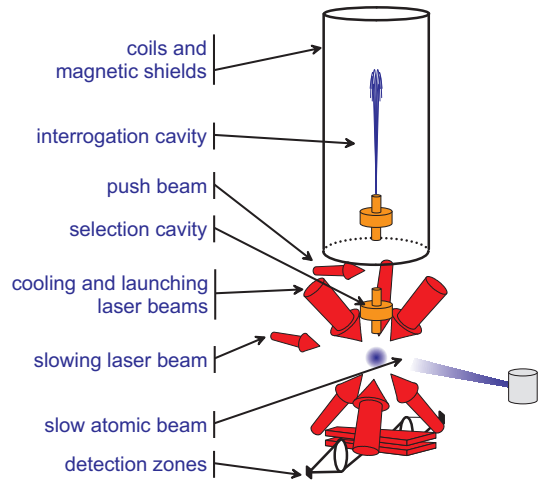}
\end{center}
\caption{Schematic view of an atomic fountain.} \label{fig:fountain}
\end{figure}

Atoms are launched upwards at 3.94~m.s$^{-1}$ by using a moving
optical molasses and cooled to $\sim 1~\mu$K in the moving frame by
adiabatically decreasing the laser intensity and increasing the
laser detuning. Atoms are then selected by means of a microwave
excitation in the selection cavity performed in a bias magnetic
field of $\sim 20$~$\mu$T, and of a push laser beam. Any of the
$|F=3,m_F\rangle$ states can be prepared with a high
degree of purity (few $10^{-3}$) by tuning the selection microwave
frequency. 52~cm above the capture zone, a cylindrical copper cavity
(TE$_{011}$ mode) is used to probe the
$|F=3,m_F\rangle\leftrightarrow
|F=4,m_F\rangle$ hyperfine transition at 9.2~GHz. The
Ramsey interrogation method is performed by letting the atomic cloud
interact with the microwave field a first time on the way up and a
second time on the way down. After the interrogation, the
populations $N_{F=4}$ and $N_{F=3}$ of the two hyperfine levels are
measured by laser induced fluorescence, leading to a determination
of the transition probability $P=N_{F=3}/(N_{F=3}+N_{F=4})$ which is
insensitive to atom number fluctuations. One complete fountain cycle
from capture to detection lasts 1045~ms in the present experiment.
From the transition probability, measured on both sides of the
central Ramsey fringe, we compute an error signal to lock the
microwave interrogation frequency to the atomic transition using a
digital servo loop. The frequency corrections are applied to a
computer controlled high resolution DDS synthesizer in the microwave
generator. These corrections are used to measure the atomic
transition frequency with respect to the local reference signal used
to synthesize the microwave frequency.

\begin{figure}[b]
\begin{center}
\includegraphics[width=8cm]{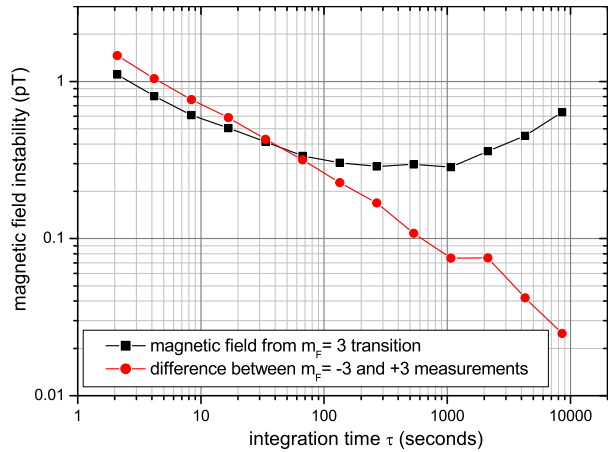}
\end{center}
\caption{Magnetic field instability as a function of integration
time $\tau$.} \label{fig:field_stability}
\end{figure}

The homogeneity and the stability of the magnetic field in the
interrogation region is a crucial point for the experiment. A
magnetic field of $200$~nT is produced by a main solenoid (length
815~mm, diameter 220~mm) and a set of 4 compensation coils.
These coils are surrounded by a first layer of 3 cylindrical
magnetic shields. A second layer is composed of 2 magnetic shields
surrounding the entire experiment (optical molasses and detection
zone included). Between the two layers, the magnetic field
fluctuations are sensed with a flux-gate magnetometer and stabilized
by acting on 4 hexagonal coils. The magnetic field in the
interrogation region is probed using the
$|F=3,m_F=1\rangle\leftrightarrow
|F=4,m_F=1\rangle$ atomic transition with a sensitivity
of $7.0084$~Hz.nT$^{-1}$. Measurements of the transition frequency
as a function of the launch height show a peak to peak spatial
variation of $230$~pT over a range of 320~mm above the interrogation
cavity. Measurements of the same transition as a function of time at
the launch height of 791~mm show a magnetic field instability near
2~pT at $\tau=$1~s as indicated in figure \ref{fig:field_stability}.
The long term behavior exhibits residual variations of the magnetic
field induced by temperature fluctuations which could cause
variations of the current flowing through solenoid, of the solenoid
geometry, of residual thermoelectric currents, of the magnetic
shield permeability, etc.

The experimental sequence is tailored to circumvent the limitation
that the long term magnetic field fluctuations could cause. First
$|F=3,m_F=-3\rangle$ atoms are selected and the
$|F=3,m_F=-3\rangle\leftrightarrow
|F=4,m_F=-3\rangle$ transition is probed at half maximum
on the red side of the resonance (0.528~Hz below the resonance
center). The next fountain cycle, $|F=3,m_F=+3\rangle$
atoms are selected and the
$|F=3,m_F=+3\rangle\leftrightarrow
|F=4,m_F=+3\rangle$ transition is also probed at half
maximum on the red side of the resonance. The third fountain cycle,
$|F=3,m_F=-3\rangle$ atoms are selected and the
$|F=3,m_F=-3\rangle\leftrightarrow
|F=4,m_F=-3\rangle$ transition is probed at half maximum
on the blue side of the resonance (0.528~Hz above the resonance
center). The fourth fountain cycle, $|F=3,m_F=+3\rangle$
atoms are selected and the
$|F=3,m_F=+3\rangle\leftrightarrow
|F=4,m_F=+3\rangle$ transition is probed on the blue side
of the resonance. This 4180~ms long sequence is repeated so as to
implement two interleaved digital servo loops finding the line
centers of both the $|F=3,m_F=-3\rangle\leftrightarrow
|F=4,m_F=-3\rangle$ and the
$|F=3,m_F=+3\rangle\leftrightarrow
|F=4,m_F=+3\rangle$ transitions. With this method,
magnetic field fluctuations over timescales longer than 4~s are
filtered in the comparison between the two transition frequencies.
Every 400 fountain cycles, the above sequence is interrupted and the
regular clock transition
$|F=3,m_F=0\rangle\leftrightarrow
|F=4,m_F=0\rangle$ is measured for 10~s allowing for an
absolute calibration of the local frequency reference with a
suitable statistical uncertainty. The overall statistical
uncertainty of the experiment is dominated by the short term
($\tau\leq 4$~s) magnetic field fluctuations (fig.
\ref{fig:field_stability}).

\begin{figure}[b]
\begin{center}
\includegraphics[width=10cm]{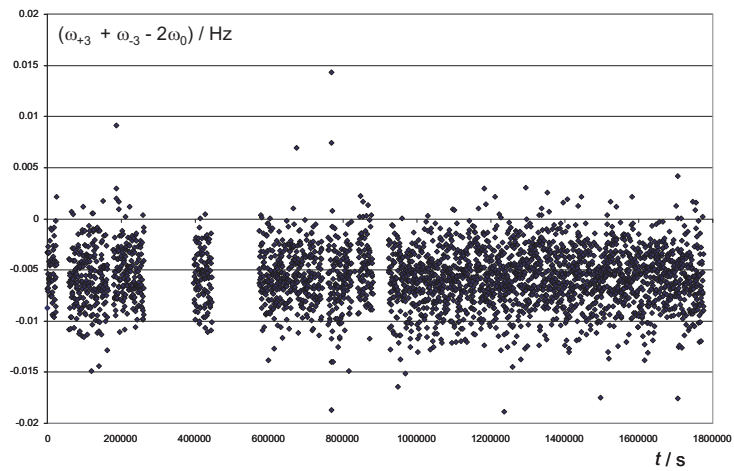}
\end{center}
\caption[]{Raw data of the measurements of $(\omega_{+3}+\omega_{-3}-2\omega_{0})$ spanning $\approx 21$ days.}
\label{SMEclock1}
\end{figure}

\begin{figure}[b]
\begin{center}
\includegraphics[width=10cm]{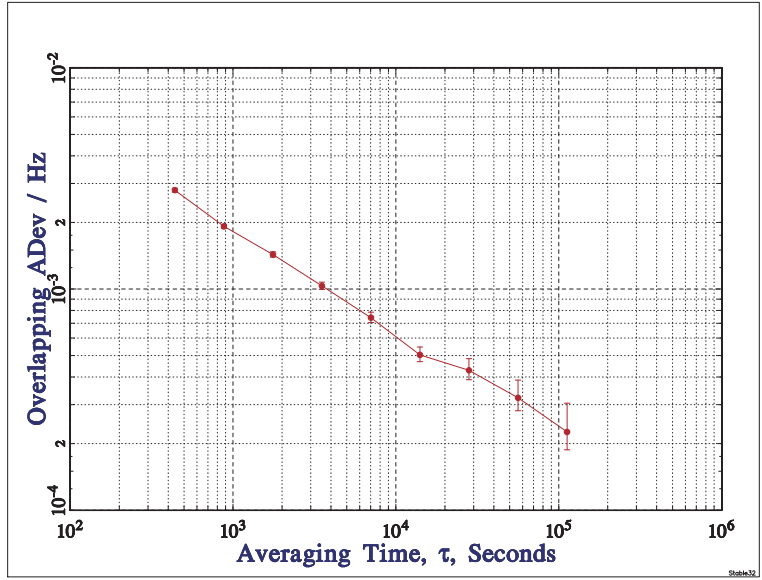}
\end{center}
\caption[]{Frequency stability of the last $\approx 10$ days of the data in figure \ref{SMEclock1}.}
\label{SMEclock2}
\end{figure}

We have taken data implementing the experimental sequence described above over a period of 21 days starting on march 30, 2005. The complete raw data (no post-treatment) is shown in figure \ref{SMEclock1}, each point representing a $\approx$432 s measurement sequence of $\omega_{+3}+\omega_{-3}-2\omega_{0}$ as described above. Figure \ref{SMEclock2} shows the frequency stability of the last continuous stretch of data ($\approx$10 days). We note the essentially white noise behavior of the data on figure \ref{SMEclock2}, indicating that the experimental sequence successfully rejects all long term variations of the magnetic field or of other perturbing effects.

According to equation (\ref{totalshift}) the frequency of the observable should be the sum of the putative Lorentz violating signal and of the second order Zeeman and other possible systematic corrections. Figure \ref{SMEclock1} shows a clear offset of the data from zero, which, using a least squares fit, is found to be $(-5.5\pm 0.1)$ mHz with a very slight linear drift of $(-1.8\pm 1.0)\times 10^{-7}$ mHz s$^{-1}$.

For our magnetic field of 202.65 nT the second order Zeeman correction of the $\omega_{+3}+\omega_{-3}-2\omega_{0}$ observable is $-2.0$ mHz. This only partly explains the offset observed in the data. The remaining part is most likely due to the differential influence of the magnetic field on the $m_F=\pm 3$ transitions, resulting from slightly different trajectories of the atoms in the different $m_F$ states and magnetic field inhomogeneities (residual first order Zeeman effect). Such differences in the trajectories could be due to differences in the trapping and/or launching of the atoms, related to the slightly different response of the Zeeman substates to the trapping fields. To check this hypothesis we have looked at the time of flight (TOF) of the atoms as a function of $m_F$. An offset of  $\approx150 \mu$s  between the $m_F=+3$ and $m_F=-3$ TOF is observed. We are presently studying this effect in more detail (Monte Carlo simulations using the magnetic field map, tests with $m_F=\pm 1$ and $m_F=\pm 2$ states, longer term observation of the TOF difference and its variation, etc.) in order to be able to completely characterize its influence on the offset in figure \ref{SMEclock1}, and its variation at sidereal and semi-sidereal frequencies.

In this paper we provide, as a preliminary results, only the values and statistical uncertainties of the coefficients $C_{\omega_\oplus}$, $S_{\omega_\oplus}$, $C_{2\omega_\oplus}$, and $S_{2\omega_\oplus}$ obtained from a model of the form

\begin{eqnarray}
\label{model}
\frac{1}{2 \pi}(\omega_{+3}+\omega_{-3}-2\omega_{0})=A~t+B&+&C_{\omega_\oplus}{\rm cos}(\omega_\oplus t)+S_{\omega_\oplus}{\rm sin}(\omega_\oplus t) \\
&+&C_{2\omega_\oplus}{\rm cos}(2\omega_\oplus t)+S_{2\omega_\oplus}{\rm sin}(2\omega_\oplus t), \nonumber
\end{eqnarray}
and the corresponding order of magnitude limits we expect for the $\tilde{c}^p$ parameters (cf. equations (\ref{obsshift}), (\ref{c_transform})) of the SME.

\begin{figure}[b]
\begin{center}
\includegraphics[width=10cm]{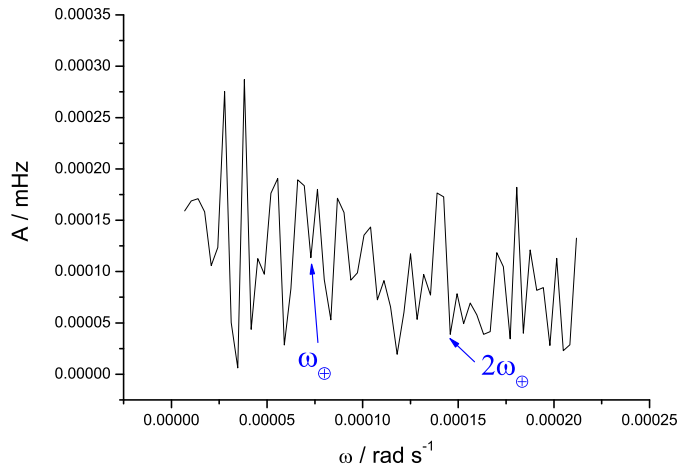}
\end{center}
\caption[]{Fitted Amplitudes $A_\omega$ for a range of frequencies around the frequencies of interest (indicated by arrows).}
\label{SMEclock3}
\end{figure}

Figure \ref{SMEclock3} shows the amplitudes $A_\omega = \sqrt{C_\omega^2+S_\omega^2}$ of least squares fits for a range of frequencies including the two frequencies of interest. We note no particularly significant peak at any frequency, and even less so at the frequencies of interest. A least squares fit at those frequencies yields the results shown in table \ref{fitresults}. The correlation coefficients between any two of the four parameters in table \ref{fitresults} do not exceed 0.07.

\begin{table}
\caption{\label{fitresults}Results of the least squares fit of equation (\ref{model}) to our complete data. Units are $10^{-5}$ Hz.}
\begin{center}
\renewcommand{\arraystretch}{1.4}
\setlength\tabcolsep{5pt}
\begin{tabular}{cccc}
\hline\hline
$C_{\omega_\oplus}$&$S_{\omega_\oplus}$&$C_{2\omega_\oplus}$&$S_{2\omega_\oplus}$ \\
\hline
$-5.3\pm 7.3$&$-10.1\pm 7.2$&$-3.2\pm 7.2$&$2.7\pm 7.2$ \\
\hline\hline
\end{tabular}
\end{center}
\end{table}

From equations (\ref{obsshift}), (\ref{c_transform}) and table I of \cite{Bluhm} we deduce orders of magnitude for the limits on the $\tilde{c}^p$ parameters of the SME (see table \ref{SME_tab}). We expect to obtain limits on two combinations of the five parameters $\tilde{c}_X^p$, $\tilde{c}_Y^p$, $\tilde{c}_Z^p$, $\tilde{c}_Q^p$, $\tilde{c}_-^p$ at a level of $10^{-25}$ GeV, and two combinations of the three parameters $\tilde{c}_{TX}^p$, $\tilde{c}_{TY}^p$, $\tilde{c}_{TZ}^p$ at a level of $10^{-19}$ GeV.

In summary, we have carried out an experiment using Zeeman transitions in a cold atom $^{133}$Cs fountain clock to test Lorentz invariance in the framework of the matter sector of the SME. In this paper we give a detailed description of the experiment and the theoretical model, we show our data and statistics, and we discuss our still ongoing investigation of systematic effects. Pending the outcome of that investigation and a more detailed theoretical analysis of our experimental results (explicit transformation of $\tilde{c}_q^p$ for our case), we provide only first estimates of the limits that our experiment can set on linear combinations of 8 SME matter sector parameters for the proton. These limits would correspond to first ever measurements of some parameters, and improvements by 11 and 14 orders of magnitude on others. A complete analysis (including systematics) of our experiment with final results for the SME parameters and their uncertainties will be the subject of a near future publication.

\section{Conclusion}

One hundred years after the publication of Einstein's original paper \cite{Einstein1905} special relativity, and its fundamental postulate of Lorentz invariance (LLI) are still as "healthy" as in their first years, in spite of theoretical work (unification theories) that hint towards a violation of LLI, and tremendous experimental efforts to find such a violation. Our experiments over the last years have provided some of the most stringent tests of LLI \cite{WP,Wolf2003,WolfGRG,Wolf2004}, but have nonetheless only joined the growing number of experiments in scientific history that measure zero deviation from LLI, albeit with an ever decreasing uncertainty. In spite of that, experimental tests of LLI are continuing along two lines: decrease of the uncertainties (see for example the contributions on rotating Michelson-Morley experiments in this volume) on one hand, and new types of experiments, e.g. the atomic clock test reported here, on the other.

In this paper we have presented a review of our recent Michelson-Morley and Kennedy-Thorndike experiment (section \ref{Molly}), and reported first results of our ongoing experiment that tests Lorentz invariance in the matter sector using a cold Cs atomic fountain clock (section \ref{FO2}). We have briefly described the two theoretical frameworks used to model and analyze our experiments (the Robertson-Mansouri-Sexl (RMS) framework and the standard model extension (SME)), and derived experimental limits on a number of parameters of those frameworks. When compared to other experiments those limits are the most stringent at present for several parameters (see tables \ref{MStab}, \ref{SMEphoton}, \ref{SME_tab}).

The next generation of Michelson-Morley experiments are based on similar technology as our experiment (section \ref{Molly}) or the equivalent approach at optical frequencies \cite{Muller}, but take advantage of active rotation of the experiment (see the corresponding contributions in this volume). Rotation of the experiment (typically at about 0.1 Hz) allows much faster data integration and places the signal modulation frequency close to the optimum where resonators are the most stable. It is expected that such experiments will lead to order(s) of magnitude improvements on orientation dependent parameters in the theoretical frameworks, but they present no advantage for only velocity dependent parameters. For example, in the RMS rotating experiments are likely to provide new, more stringent limits for the Michelson-Morley parameter ($P_{MM}=1/2-\beta_{\mathrm{MS}} +\delta_{\mathrm{MS}}$) but no improvements on the Kennedy-Thorndike one ($P_{KT}=\beta_{\mathrm{MS}} -\alpha_{\mathrm{MS}} -1$). So we expect our (and other) present limits on $P_{MM}$ to be significantly improved, but we see no obvious way of improving on our present limit on $P_{KT}$ in the near future. 

Several improvements of our clock test of LLI in the SME matter sector (section \ref{FO2}) are possible. For example, using the unique capability of our double fountain (FO2) to run on both, Cs and Rb, we expect to be able to use Cs as the SME sensitive species and Rb (which is less sensitive to the SME \cite{Bluhm}) as the magnetic field probe. In that way we should be able to perform magnetic field independent measurements that could improve on our present results, and allow access to other SME parameters that we are insensitive to with our present set up. Also, rotation of the experiment could provide a method for faster modulation of the signal but is unpractical in an Earth bound laboratory. However, space missions with onboard atomic clocks are well suited for such a test. In particular the European ACES (Atomic Clock Ensemble in Space) mission \cite{ACES}, scheduled for flight on the international space station (ISS) in 2009, seems very promising in this respect. It will include a laser cooled  Cs clock (PHARAO) with expected performance at least equivalent to our FO2, but with the orientation of its quantization field axis modulated at a 90 min period (ISS orbital period) rather than 24 hr as in our case. This should allow for much faster data integration and significant improvement on the limits presented here.


\begin{thebibliography}{99}
\bibitem{masercom} We assume here that local position invariance is sufficiently verified so that the variation of the maser frequency due to the diurnal variation of the local gravitational potential is negligible. Indeed the results of \cite{Bauch} imply that such variations should not exceed 2 parts in $10^{-17}$ which is significantly below our noise level.
\end{thebibliography}
\end{document}